\documentclass[aps,pra,twocolumn,showpacs,superscriptaddress,10pt]{revtex4-1}

\usepackage{amsmath}
\usepackage{amsfonts}
\usepackage{amssymb}
\usepackage{graphicx}
\usepackage{color}
\usepackage{soul}
\usepackage{xcolor}
\usepackage{footmisc}
\usepackage{braket}
\allowdisplaybreaks

\begin{document}
\title{ 
Dicke time crystals in driven-dissipative quantum many-body systems }

\author{B. Zhu}
\affiliation{ITAMP, Harvard-Smithsonian Center for Astrophysics, Cambridge, MA 02138, USA}
\affiliation{Department of Physics, Harvard University, Cambridge MA, 02138, USA}
\author{J. Marino}
\affiliation{Department of Physics, Harvard University, Cambridge MA, 02138, USA}
\affiliation{Department of Quantum Matter Physics, University of Geneva, 1211, Geneve, Switzerland}
\affiliation{Kavli Institute for Theoretical Physics, University of California, Santa Barbara, CA 93106-4030, USA}
\author{N. Y. Yao}
\affiliation{Department of Physics, University of California, Berkeley, CA 94720, USA}
\affiliation{Materials Sciences Division, Lawrence Berkeley National Laboratory, Berkeley CA 94720, USA}
\author{M. D. Lukin}
\affiliation{Department of Physics, Harvard University, Cambridge MA, 02138, USA}
\author{E. A. Demler}
\affiliation{Department of Physics, Harvard University, Cambridge MA, 02138, USA}

\begin{abstract}
The Dicke model -- a paradigmatic example of superradiance in quantum optics -- describes an ensemble of atoms which are collectively coupled to a leaky cavity mode. 
As a result of the cooperative nature of these interactions, the system's dynamics are captured by the behaviour of a single mean-field, collective spin.
In this mean-field limit, it has recently been shown that the interplay between photon losses and periodic driving of light-matter coupling can lead to time-crystalline-like behaviour of the collective spin \cite{Gong}. 
In this work, we investigate whether  such a Dicke time crystal is stable to perturbations that explicitly break the mean-field solvability of the conventional Dicke model.
In particular, we consider the addition of short-range interactions between the atoms which breaks the collective coupling and leads to complex many-body dynamics. 
In this context, the interplay between periodic driving, dissipation and interactions yields a rich set of dynamical responses including long-lived and metastable Dicke time crystals, where losses can cool down the many-body heating resulting from the continuous pump of energy from the periodic drive. 
Specifically, when the additional short-range interactions are ferromagnetic, we observe time crystalline behaviour at non-perturbative values of the coupling strength, suggesting the possible existence of stable dynamical order in a driven-dissipative quantum many-body system.
These findings illustrate the rich nature of  novel dynamical responses with many-body character in quantum optics platforms.%
%

\end{abstract}

\date{\today}
\maketitle

\section{Introduction }
The  study of emergent dynamical phenomena in interacting quantum many-body systems constitutes a frontier of research in modern quantum optics and condensed matter physics. 
In this quest for phases of quantum matter without equilibrium counterpart, time crystals (TC) represent a promising  candidate for a novel form of dynamical order  out-of-equilibrium. In TCs,  observables dynamically entrain 
 at a  frequency subharmonic of the one imposed by an external periodic drive~\cite{saph, Wilczek, Bruno3,Noz, Sacha, Wantabe,Else, Khe,  Key, Key2, Khe2, Yao,yao2,Gong, Fazio, noi, marcuzzi}, and they have been currently realised with  trapped ions~\cite{Monroe} and solid state systems~\cite{Lukin,seanPRL2018,seanPRB2018}. In most previous studies, TCs are  realised in closed interacting quantum many-body systems, which are  prone to heating towards an infinite temperature state under the  action of periodic drive~\cite{dasarn, dal}, therefore, a slowdown of energy absorption is customarily entailed via a disorder induced many-body localized phase~\cite{ponte, ponte2, zhang, bordia}, or by fast driving~\cite{aba, PhysRevLett.116.120401, Else, Ho, ElsePRX}.

An alternative pathway could consist in ``cooling'' time
crystals via coupling to a cold bath, which can absorb the energy pumped by the periodic drive~\cite{ElsePRX}. A natural candidate to explore this avenue is represented by a recent  line of inquiry on the exploration of TC-like behaviour in the open Dicke model, which describes an ensemble of atoms collectively coupled to a   leaky photon cavity mode. 
The periodic drive of Dicke light-matter interactions in the superradiant regime can entail   sub-harmonic dynamical responses~\cite{Gong},
however, the collective nature of  interactions renders the dynamics of this class of TCs equivalent to a single body problem consisting of a mean-field  collective  spin degree of freedom moving on the Bloch sphere. Our key goal is to understand the stability of the Dicke time crystal when one breaks the mean-field nature of the model. 
\begin{figure}[t]
  \includegraphics[width=0.4\textwidth]{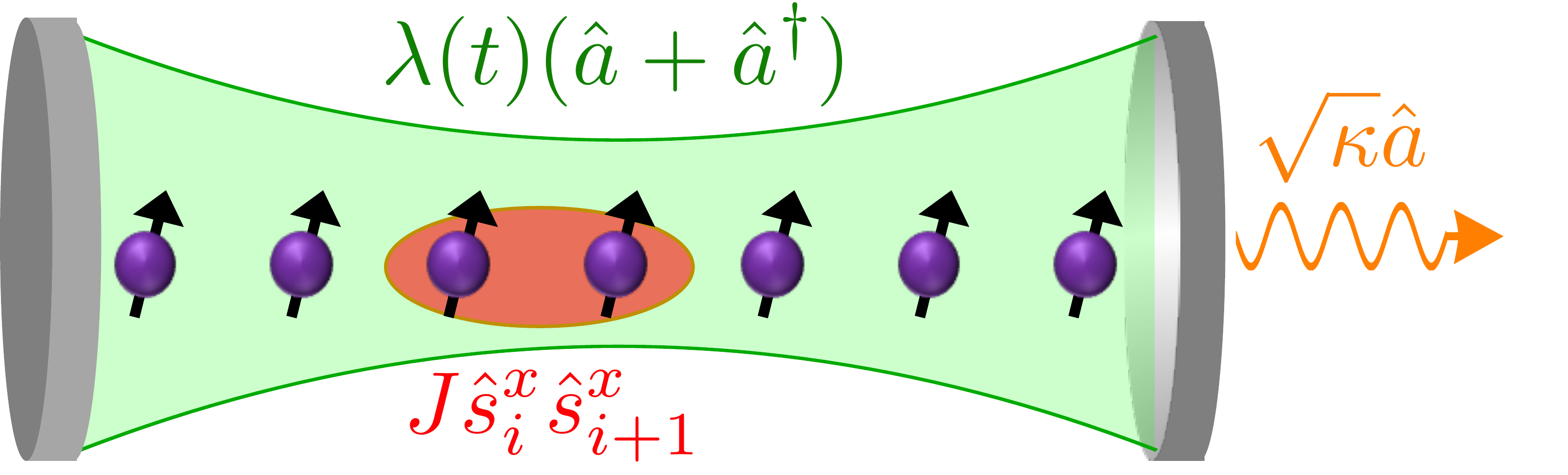}
  \caption{The driven-dissipative many-body Dicke model studied in this work. An ensemble of atoms are collectively coupled to a  photon field $\hat a$ with a time-varying strength $\lambda(t)$, which consistutes a  Floquet driving that injects energy into the system. The atoms also interact with each other via a short-range interaction of strength $J$, which breaks the collective nature of the system. Dissipation of energy into a bath is included through the loss of photons at a rate $\kappa$.  }
\label{fig1}
\end{figure}

\begin{figure*}[t]
\includegraphics[width=0.6\textwidth]{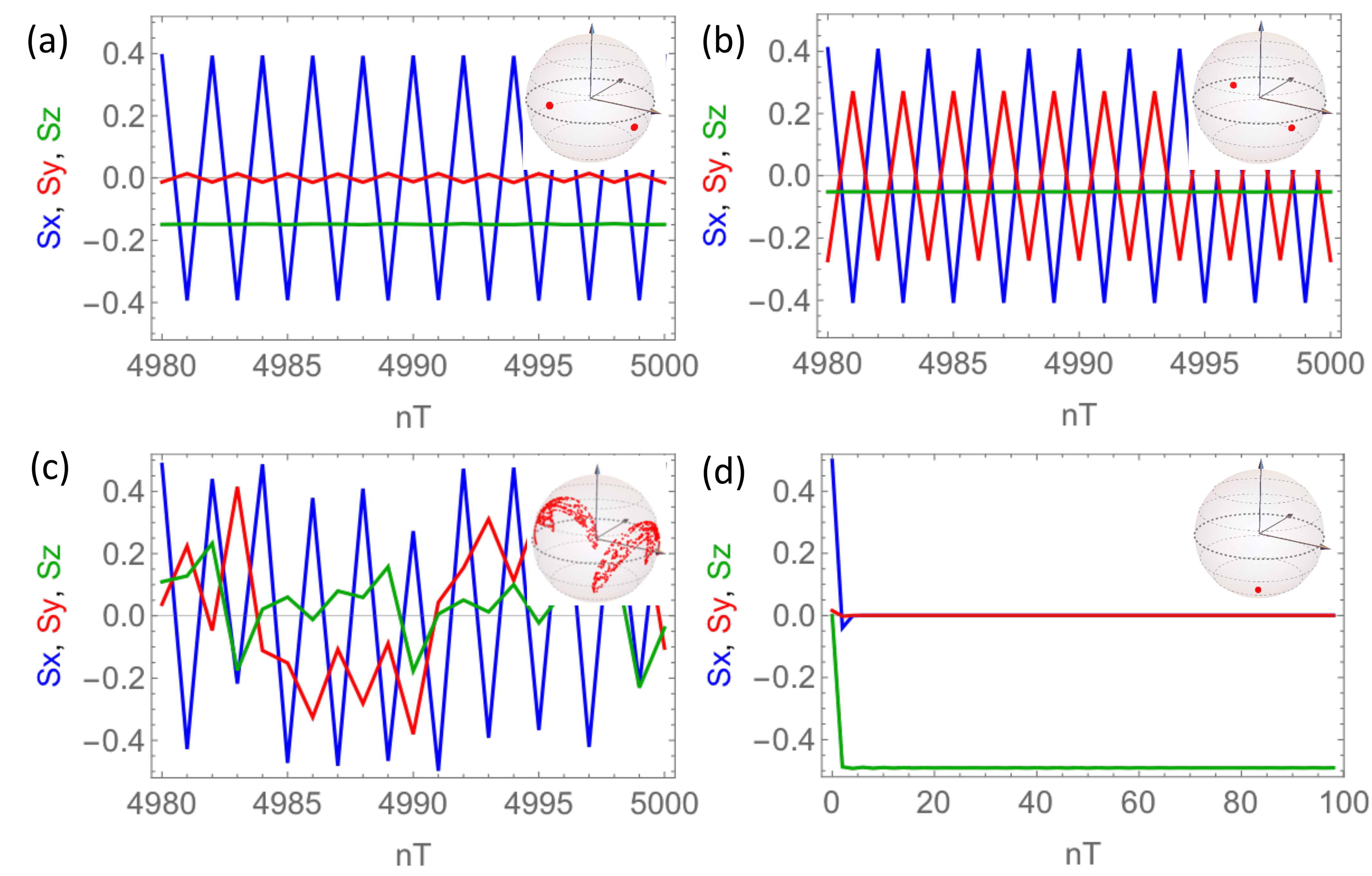}
\caption{Instances of the stroboscopic dynamics  of the collective spin projections, $S_x$, $S_y$, and $S_z$ (colored in blue, red and green respectively and each normalized by atom number $N$), for  Dicke-TC (a and b), IR (c), and OD (d) dynamical responses. (a) $J/\Omega$=0.056, $\kappa/\Omega=0.375$. (b) $J/\Omega$=0.2, $\kappa/\Omega=1.45$. (c) $J/\Omega$=0, $\kappa/\Omega=0.9$.
  (d) $J/\Omega$=0.04, $\kappa/\Omega=3.2$. The insets show the stroboscopic snapshots on the Bloch sphere for the last 1000 periods. 
 }
\label{fig2}
\end{figure*}
To this end, we explore the robustness of Dicke-TCs to local interactions which break the collective coupling of the original model (Fig.~\ref{fig1}). 
We observe that this class of Dicke-TCs can remain stable to such mean-field breaking perturbations in certain limits. Crucially, this lifts the phenomenon from an inherently collective, mean-field  effect to the steady-state behaviour of a dissipative many-body system. We note however, that unlike the traditional venue for discrete time crystals \cite{Khe,  Key, Key2, Khe2, Yao}, where short-range interactions are essential for \emph{stabilizing} time crystalline order, here, the short-range interactions are rather viewed as \emph{perturbations} to the original mean-field Dicke-TC.

The interplay between Floquet driving, dissipation, and interactions results in a rich set of dynamical responses. In particular, we find regimes where TCs are stabilised by the bath, which counteracts the energy pumped into the system by the drive. We also observe the emergence of metastable dissipative TCs, characterised by a  slowly  decaying envelope evolving eventually  into a trivial steady state dominated by dissipation. In addition, we find a family of ferromagnetic driven-dissipative TCs with strong resilience to many-body heating.  

\section{The model}
We consider an chain of $N$ two-level atoms with short-range interactions among each other
\begin{equation}\label{int}
  \hat H_{int}=J\sum^N_{i=1}\hat s_i^x\hat s_{i+1}^x,
\end{equation}
where $\hat s^{x,y,z}=\hat\sigma^{x,y,z}/2$, and $\hat \sigma^{x,y,z}$ are Pauli matrices. The atoms are collectively coupled to a photon field, {\em e.g.}, by placing them inside an optical cavity~(Fig.~1), which can be described by the hamiltonian~\cite{Car,ritschReview}
\begin{equation}\label{eq:ham}
  \hat H_{ac}=\omega \hat a^\dagger\hat a+\omega_0\hat S_z+\frac{2\lambda(t)}{\sqrt{N}}(\hat a+\hat a^\dagger)\hat S_x,
\end{equation}
where $\hat S_{x,y,z}=\sum_i^N\hat s_i^{x,y,z}$. We allow the light-matter coupling to be  varied in time, $\lambda(t)$. Dissipation occurs when  photons leak out of the cavity, as encoded by the quantum master equation
\begin{equation}\label{eq:master}
\dot{\hat\rho}=-i[\hat H,\hat\rho]+\frac{\kappa}{2}(2\hat a\hat\rho\hat a^\dagger-\hat a^\dagger \hat a\hat \rho-\hat\rho\hat a^\dagger \hat a),
\end{equation}
for the total density matrix of the system, $\hat{\rho}$, where $\hat H=\hat H_{ac}+\hat H_{int}$, and $\kappa$ characterises the rate of photon loss.

When $J=0$, the above reduces to the well-known open Dicke model~\cite{HeppLieb, wang, nard, duncan, garr,nagy, nagy11, PhysRevLett.105.043001, Bhaseen2012, DallaTorre2013, lang2016critical, PhysRevA.94.061802, Gelhausen2017, yulia}. As the coupling is only between the single photon mode and the  collective spin operator, $\hat{S}^x$, the   Dicke model is exactly  solvable in the thermodynamic limit $N\to\infty$: its dynamics can be described by the  mean-field  motion of the photonic amplitude, $\langle a \rangle $, coupled to three  classical degrees of freedom,   $\langle \hat S_{x,y,z} (t) \rangle$, evolving on the Bloch sphere. When $J\neq 0$, short-range atom-atom interactions break the exact solvability  of $H_{ac}$, spoiling the collective character of  the Dicke hamiltonian. In addition to the collective mode $\vec S$, which corresponds to the $k=0$ Fourier mode $\vec{s}_k\equiv\sum^N_{j=1} e^{-ikj}\vec{s_j}$, all other $k\neq 0$ modes could also be excited.  Hence $H_{int}$ introduces quantum fluctuations in the spin (or atomic) degrees of freedom, which  require treating the dynamics in Eq.~\eqref{eq:master} as a  quantum many-body problem.

We simultaneously account for dissipation and quantum fluctuations using   a time-dependent spin-wave approach, which has been demonstrated  effective in capturing dynamical quantum many-body effects~\cite{jamir, jamir3,jamir2}. Specifically, we first perform a time-dependent rotating frame transformation to  align the time-dependent $\hat{z}$-axis with the collective Bloch vector $\vec{S}(t)$, and then in this co-moving `frame' we perform a Holstein-Primakoff transformation in order to expand  the spin operators in $H_{int}$  to the lowest-order in  the density of spin-wave excitations, $\epsilon(t)$ (see Appendix~\ref{app:spinwave} for details). The many-body effects introduced by $H_{int}$ are encoded in the dynamical coupling between spin-waves, and the collective spin as well as photon field. Excitation of spin waves leads to a depletion of the $k=0$ mode, $2|\vec{S}(t)|/N=1-\epsilon(t)$, similarly in spirit  to approaches which incorporate self-consistently the effect of quantum fluctuations in the dynamics of a ``condensate''~\cite{CITRO2015694}.  The spin-wave density, $\epsilon(t)$, representing the total population of all $k\neq 0$ spin-wave excitations,  is required to remain small at all times in order to have a self-consistent lowest-order Holstein-Primakoff expansion. By monitoring the growth of spin-wave density, we can identify regions of ``heating'', where  the  collective spin order shrinks under effect of strong many-body interaction, accompanied by a large value of $\epsilon(t)$ in dynamics~(see Appendix~\ref{app:spinwave}). 




\section{Dynamical responses }
We impose a periodic modulation on the light-matter coupling $\lambda(t)$: during a first ``bright-time'', $nT\leq t<(n+1/2)T$, we set  $\lambda(t)=\lambda=\Omega$, and during the ``dark-time'', $(n+1/2)T\leq t<nT$, we switch off $\lambda(t)$. Here, $\Omega=2\pi/T=(\omega_0+\omega)/2$ denotes the driving frequency. This periodic driving counteracts with the energy loss through photon leaking, and as previously found in Ref.~\cite{Gong} for the collective Dicke model ($\hat H_{int}$ absent), it entails  a period-doubling response in spin observables. The presence of such TC-like behaviour can be understood  from the $\mathbb{Z}_2$ symmetry of the Dicke model with constant $\lambda$, under the parity operator $P= e^{i\pi(\hat{a}^\dag\hat{a}+\hat{S}_z+\frac{N}{2})}$. For $\lambda>\lambda_c=\frac{1}{2}\sqrt{{(\omega^2+{\kappa^2}/{4})\omega_0}/{\omega}}$, a quantum phase transition that breaks the $\mathbb{Z}_2$ symmetry occurs, and the system enters a superradiant phase, featuring two steady states, with  spin projection $\langle S_x\rangle/N=\pm X$, $\langle S_y\rangle=0$ and non-vanishing photon amplitude $a=\mp \sqrt{N}\lambda X/(\omega-i\kappa/2)$ ($X=\frac{1}{2}\sqrt{1-\lambda_c^4/\lambda^4}$). When $\delta\equiv(\omega_0-\omega)/\Omega=0$, the free evolution during the `dark-time' accumulates a phase of $\pi$ for both $S_x=\langle S_x\rangle$ and $a$, and the system switches from one steady state to another, {\em i.e.}, $ S_x\rightarrow - S_x$ and $a\rightarrow -a$. As a result, the dynamics repeats after two cycles of the driving and  a sub-harmonic response with period $2T$ appears. In such a picture, atoms are simply described as a single classical spin, which becomes invalid in the presence of $\hat H_{int}$, and thus it deserves a careful investigation whether the Dicke-TC exists in a many-body system with nonzero $J$. 

We explore the spin dynamics for various interaction strength $J$ and dissipation rate $\kappa$, using the time-dependent spin-wave approach. Our analysis shows that the Dicke-TC order can exist beyond the collective case ($J = 0$) and survive many-body interactions. 
For a range of finite $J/\Omega\ll 1$ and  $0<\kappa/\Omega\lesssim1$, we observe a stable subharmonic response in $S_{x}(t)$, as plotted in Fig.~\ref{fig2}(a) for an instance, where the spin-wave density remains small and therefore it is robust to heating~(see Appendix~\ref{app:S1}). Upon increasing the values of $J$, the inelastic scattering induced by many-body interactions becomes efficient, provoking a sizeable growth of spin-wave density, which invalidates the lowest-order spin-wave expansion and makes the collective spin, $\vec{S}$, crumble. In this regime of strong interactions, the system is prone to  heating under the action of the Floquet driving (see also ~\footnote{In this work, we choose as upper threshold for spin wave density the value, $\epsilon\sim0.2$, in order to delimit the heating region from the other dynamical responses\label{fn:trunc}}). However, as shown in  Fig.~\ref{fig2}(b), for moderate strengths of $J/\Omega\lesssim 1$,  such effect of heating can still be remediated with sufficient dissipation $\kappa/\Omega\gtrsim1$~(see also Appendix~\ref{app:S1} and \ref{app:phase}). In this case, strong dissipation acts as a 'contractor' for the dynamics, guiding swiftly the system towards the desired non-equilibrium steady state and enables stable oscillations of $S_x$. Meanwhile, the combination of strong dissipation and driving leads to remarkable disturbance of the spin state within each period, resulting in a larger value of $S_y$ compared to the Dicke-TC at small $\kappa$ (cf. Figs.~\ref{fig2}(a) and ~\ref{fig2}(b)). Here, the enhanced fluctuation of photon field  associated with strong dissipation can induce large phase noise and destroy the Dicke-TC as well; however, for sufficiently large atom number $N$, the phase noise can still be suppressed, since the  amplitude of photon field scales as $\sim \sqrt{N}$ in the superradiant regime (see Appendix~\ref{app:spinwave}), and thus allows the observation of the Dicke-TC.


While period doubling is a fragile dynamical response in one-particle periodically driven systems, i.e. it disappears as a tiny $\delta\neq0$ is switched on, the collective ($J=0$) Dicke-TC is robust in a range of small $\delta\neq0$, thanks to the macroscopic $S_x$-order built during the superradiant ``bright-time'' and thanks to the ``contractive'' role of dissipation which guides the system towards the desired non-equilibrium steady state. Such robustness to deviations from the $\delta=0$ limit, persists upon inclusion of many-body interactions, $J\neq0$ (as an example, $\delta=-0.12$ is used for Fig.~\ref{fig2}). The persistence of many-body Dicke-TCs  on time scales much longer than $t_\kappa\sim 1/\kappa$ (see the exemplary dynamics in Fig.~\ref{fig2}(a)), indicates that they represent a long-lived phenomenon, since in the presence of dissipation,  relaxation is typically expected to occur on time scales inversely proportional to the system-bath coupling, $\kappa$. Indeed, we never observe  decay of the Dicke-TC order on the longest timescales accessible to our numerical study (e.g, see Fig.~\ref{fig2}(a) and footnote~\footnote{In principle, we cannot exclude that collapse of the order parameter might occur at much later times, and  establishing the stability to infinite times may require another approach, which is beyond the scope of this work. However, this effect, even if present, would be of no practical relevance on experimentally accessible time scales.}). The existence of Dicke-TC is also insenstive to initial conditions. As we checked in our numerical simulations,   similar responses are observed over a wide range of initial conditions (see Appendix~\ref{app:S1}).

Here, dissipation is capable of stabilising rather than destroying the Dicke-TC order in $S_x(t)$ since it acts on the photon mode and therefore collectively on spin order. On the contrary, when local losses are introduced, dissipation is intrusive and detrimental, and it destroys TCs (see for instance Ref.~\cite{laz17}). Note  that in conventional  discrete TCs, instead, the interactions  serve to lock single-spin dynamics into a stable  subharmonic response that is robust to perturbations or imperfections of the drive~\cite{Yao, Monroe}.

When the rate of dissipation is intermediate, $\kappa/\Omega\sim 1$,  we recognise a region of irregular (IR) dynamics, where the trajectory of $\vec{S}(t)$ is scattered on the Bloch sphere (Fig.~\ref{fig2}(c)). In this case, the photon amplitude is sizeably reduced, and since it  contributes to building the $S_x$-order via the light-matter coupling term $\propto\lambda$, the system does not develop a  $\langle S_x\rangle$ component sufficiently strong in order to counteract the  dephasing induced by the ``transverse field''~$\propto\omega_0 \hat{S}^z$ during the 'dark-time', and this results into a featureless dynamical response lacking of period-doubling. We also notice that in this regime a relatively small value of $J$  can lead to a proliferation of spin-wave excitations, suggesting a tendency to heating (see Appendix~\ref{app:phase}).

Complementarily, with excessively  large $\kappa/\Omega\gg 1$,  while dissipation is sufficient to  cool the system and prevents many-body heating, it also destroys Dicke-TC order: in this case, dissipation overdamps dynamics, and the collective spin of the system relaxes to a trivial  steady state where all spins point down towards the south pole of the Bloch sphere~\footnote{{$\hat S_z \gtrsim -N/2$, since part of the collective spin is dissipated in the bath of spin waves.}}  (see Fig.~\ref{fig2}(d)). This overdamped regime (OD) is thus characterised by a vanishing magnitude of the spin projection, $| S_x+i S_y|$.

The different dynamical responses discussed above are summarised in a qualitative cartoon,  Fig.~\ref{figsketch}, in Appendix~\ref{app:phase}. It is interesting to note that, with increasing $J$, we find the system tends to be overdamped with a smaller $\kappa$, which is indicated by the downwards bending of the boundary between the OD and Dicke-TC regions shown in Fig.~\ref{figsketch}. Hence  fast ``cooling'' does not always protect, but can instead destroy Dicke-TC behaviour. Nevertheless, dissipation plays a crucial role in establishing the rich phenomena in Fig.~\ref{fig1} (see Appendix~\ref{app:lmg}, where the impact of  $H_{int}$ on TCs of the Lipkin-Meshkov-Glick  model is addressed). We note that, in this work, the frequency of the drive, $\Omega$, only plays the role of an overall energy scale.

%

%
%

\begin{figure}[t!]
\includegraphics[width=0.43\textwidth]{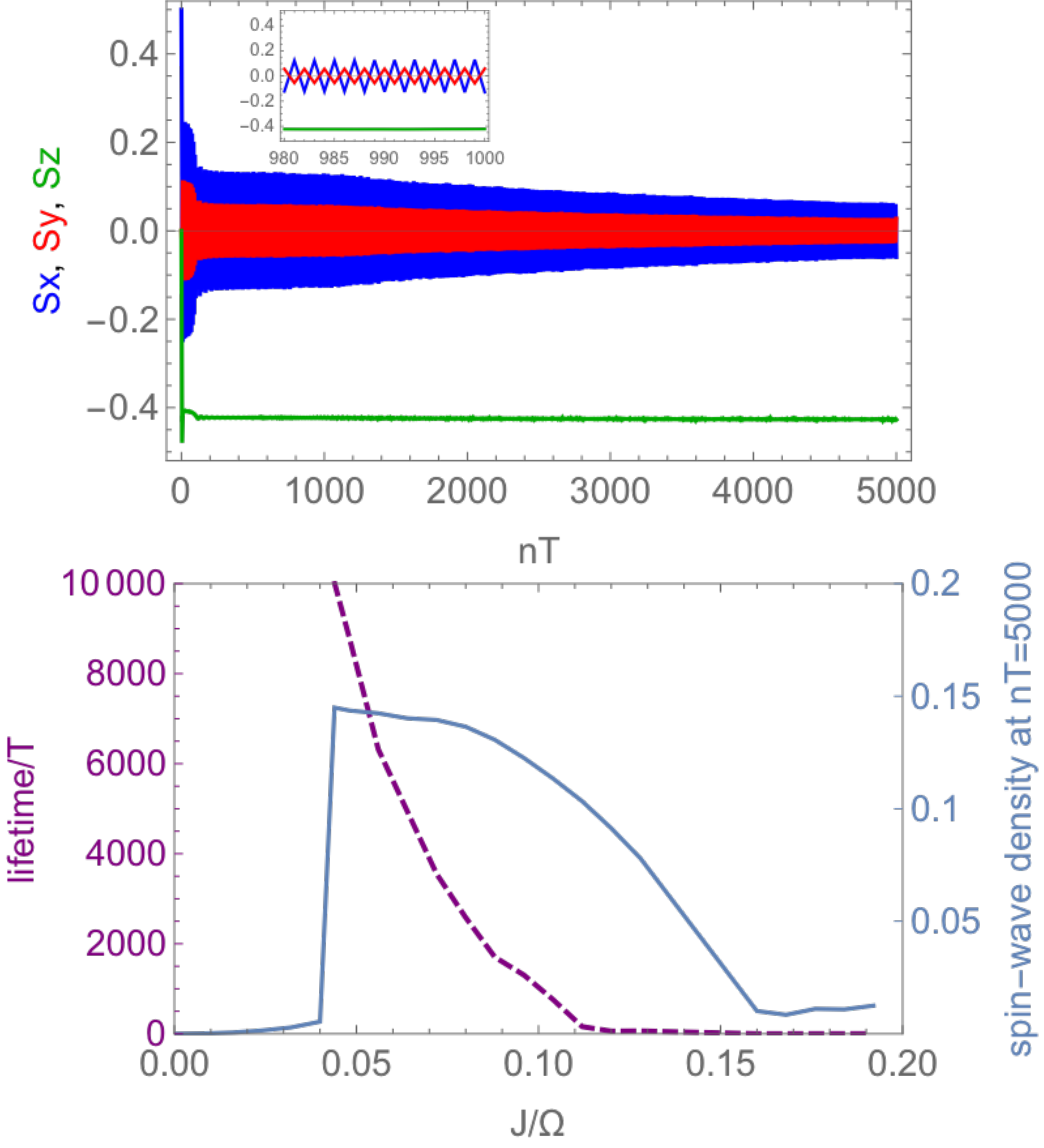}
\caption{Top: The stroboscopic dynamics of the MTC for  $\kappa/
\Omega=2.5$, $J/\Omega=0.08$, $\delta=-0.12$. The dynamics of $S_x(t)$ (blue, normalised by $N$)  appears indistinguishable from a conventional Dicke-TC response  on time windows of the order of a few decades of cycles (inset); on longer time scales it displays instead a slowly decaying envelope. Bottom: The lifetime of the MTC (defined as the time when the stroboscopic amplitude of $  S_x$ decays to $\lesssim 0.1$), and spin-wave density of the MTC after $5\times10^3$ cycles. We find that the lifetime falls  to zero following an empirical law $\propto\exp{[-A(J/\Omega)^{1.6}]}$, with $A$ a positive factor. }
\label{fig3}
\end{figure}
\section{Metastable dissipative time crystal}
For intermediate values of both $J$ and $\kappa$ (see for instance Fig.~\ref{figsketch}), our system hosts another type of nontrivial behaviour: a dissipative metastable time crystal (MTC) characterised by a slowly decaying envelope, which deteriorates, in the long time, into a trivial asymptotic state dominated by dissipation with vanishing $ S_x$ (see upper panel of Fig.~\ref{fig3}). 
This behaviour is   distinct from the Dicke-TC:  in the lower panel of Fig.~\ref{fig3} we plot the associated spin-wave density  (blue line), which exhibits 
a
discontinuous jump at a non-vanishing value of $J$, 
when the lifetime of the TC starts also to decrease~(purple line); this suggests that the
metastability is not expected to manifest for small values of  $J$, and therefore the conventional Dicke-TC represents a long-lived phenomenon (see footnote~\footnote{We have checked with a finite-size analysis  that these conclusions   do not depend on $N$.}).
The lifetime, $\tau$, of this MTC  gradually decreases with $J$, following the empirical law, $\tau\propto\exp{[-A(J/\Omega)^{1.6}]}$ (with $A$ a positive prefactor), and vanishes  when the system enters the OD regime, where also the spin-wave density  becomes  small since  the system reaches the fully polarised state in the  negative $\hat{z}$-direction of the Bloch sphere (corresponding to spin-waves vacuum).
The MTC found here appears as a genuine interplay of period driving, dissipation and interaction. We remark here that it doesn't result from  a high-frequency expansion~\cite{PhysRevLett.116.120401, aba}, and thus is distinct from  the ``Prethermal'' Floquet TCs found in previous studies~\cite{ ElsePRX, Lukin},  since $\Omega$ is  an overall energy scale in our system (see Fig.~\ref{fig1}). 
%
  A possible explanation for the phenomenon is suggested by the dynamics of $\epsilon(t)$: during the metastable evolution, the density of spin waves strongly fluctuates and is out of phase with the dynamics of the photon amplitude, accumulating at every cycle a tiny dephasing, which eventually leads  $ S_x(t)$ to collapse. 
%

  \section{Ferromagnetic driven-dissipative time crystal }
\begin{figure}[t!]
\includegraphics[width=0.4\textwidth]{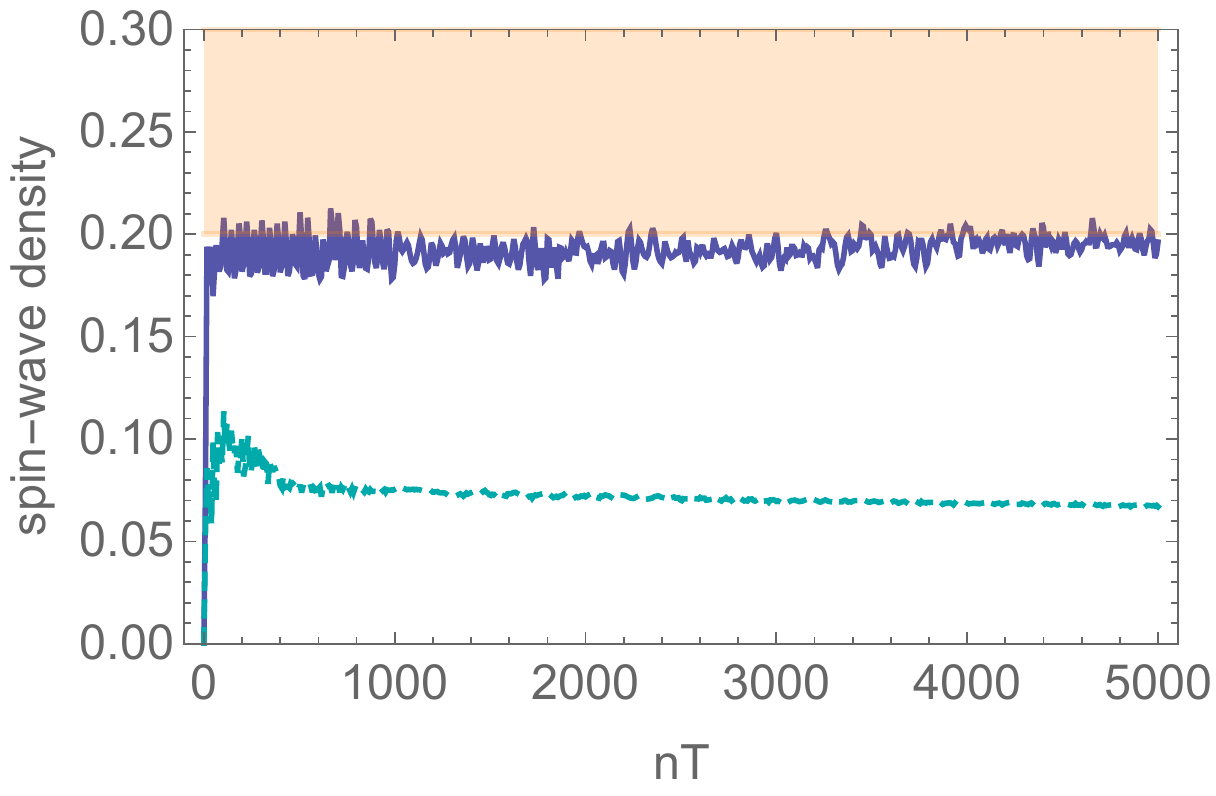}
\caption{  Comparison between the stroboscopic dynamics of spin-wave densities of a ferromagnetic (cyan dashed line, $J/\Omega=-2.4$) and  of an anti-ferromagnetic (blue line, $J/\Omega=0.2$) Dicke-TC for $\delta=-0.12$ and $\kappa/\Omega=0.4$. The orange area denotes the region of values of spin-wave density where the lowest order Holstein-Primakoff breaks and the system becomes prone to many-body heating (see footnote~[53]). }
\label{fig4}
\end{figure}
  
As discussed above, the steady states underlying the Dicke-TC order possesses a ferromagnetic nature.  When the many-body interactions $H_{int}$ is also ferromagnetic ($J<0$),  inter-spin interactions  can reinforce the  ordering along the $\pm\hat{x}$-direction, giving rise to robustness against heating and the overdamping caused by dissipation. Indeed, for non-perturbative values of $|J|\sim O(1)$, we find that a ferromagnetic ($J<0$) dissipative Dicke-TC can be stabilised at intermediate dissipation rate  without significantly heating up the system (see footnote~\footnote{In this case, we switch off the $H_{int}$ term during the ``dark-time'', in order to avoid inhomogeneous dephasing resulting from inter-spin interactions. }). This is shown in Fig.~\ref{fig4}, where we plot the spin-wave density  for anti-($J>0$) and ferro-($J<0$) magnetic TCs: a sizeable spin-wave density denotes fragility  to many-body interactions and a dynamics prone to heating, and such effects are expected to be pronounced at large $J$. Remarkably, this does not occur for ferromagnetic inter-spin interactions which develop tiny values of  $\epsilon(t)$ even for $|J|\sim O(1)$. The emergence of such a  Dicke-TC response within the coexistence of significant many-body interactions and dissipation rate, appears to us a strong incarnation of TC-like behaviour in driven-dissipative platforms: it is a novel form of dynamical order out-of-equilibrium, which significantly departs both from the mean-field Dicke TC response (where $J\simeq 0$), and from conventional (many-body) discrete TCs where dissipation is not a constitutive ingredient ($\kappa=0$). In this perspective, such ferromagnetic  TC  represents a non-equilibrium state of strongly coupled, driven-dissipative, quantum matter exhibiting rich dynamics that can trigger motivation towards the search of other non-trivial dynamical phases  in many-body quantum optics.

\section{Summary and outlook }
The Dicke model is currently engineered in several experimental platforms~\cite{black2003observation, baumann2010dicke, Baumann2011, brennecke2013real, klinder2015observation, klinder2015dynamical}. We expect  our results to be qualitatively insensitive to the details of the microscopic structure of  the interaction term $H_{int}$, and  to hold in a broader set of models, and thus would be relevant for experiments where collectivity of the system is inevitably broken by inhomogeneous fields, or  spatially varying light-matter couplings, or genuine inter-particle interactions such as using Rydberg atoms~\cite{norcia,Vaidya,gobanNatComm2014,pfaufiber2014}. 
 The stabilisation of Dicke-TC order seen for strong ferromagnetic spin-spin interactions can also be generalised to systems where the inter-spin interactions have a anti-ferromagnetic character ($J>0$), given the capability to control atom-light coupling in cavity experiments~\cite{rempegatePRX2018}. For coupling $\propto \lambda (\hat a+\hat a^\dag)\sum_j(-1)^j \hat\sigma_j^x$, a similar Dicke-TC response exists in this case but with  anti-ferromagnetic ordering. Hence a $\hat H_{int}$ with certain $J>0$ would be expected to extend the Dicke-TC to a many-body regime. Another interesting  possibility offered by the control of light-matter coupling consists in realising Dicke-TC responses with {higher integer periods ($nT$ with $n>2$)}  without employing high spin atoms (see for instance Ref.~\cite{Lukin}). In the Appendix \ref{app:quad} we show that a Dicke model with coupling of the form $\propto \lambda (\hat a+\hat a^\dag)\sum_j[(-1)^{j/2}\hat\sigma^+_j+h.c.]$  realises a dynamical response with  quadruple  period.

The setup analysed in our work is closely related to  quantum optics platform, but can also be relevant for condensed matter and solid state platforms, where the system can be modeled as a quantum spin chain coupled to a phonon bath.
We believe that the outreach of our results has the potential to motivate a new generation of experiments on  TCs in many-body systems, where the presence of a bath plays a crucial role at variance with current realisations~\cite{Lukin, Monroe}.

%

%

\section{Acknowledgements } The authors acknowledge discussions with  D. Abanin, S. Gopalakrishnan, W. W. Ho, R. Moessner, C. Nayak, V. Oganesyan,    M. Schleier-Smith, D. Sels and S. Yelin. The authors would particularly like to thank F. Machado for a careful reading of the draft and many helpful comments. 
JM acknowledges A. Lerose, A. Gambassi,  A. Silva and B. Zunkovic for previous collaborations on the subject of time-dependent spin-wave expansions.
This work is supported by the DARPA DRINQS program (award D18AC00014), the NSF (PHY-1748958),   the AFOSR-MURI Photonic Quantum Matter (award FA95501610323), the Harvard-MIT CUA, the National Science Foundation (NSF), the David and Lucille Packard Foundation, and the Vannevar Bush Faculty Fellowship.
BZ is supported by the NSF through a grant for the Institute for Theoretical Atomic, Molecular, and Optical Physics at Harvard University and the Smithsonian Astrophysical Observatory. JM is supported by the European Union's Framework Programme for Research and Innovation Horizon 2020 under the Marie Sklodowska-Curie Grant Agreement No. 745608~(`QUAKE4PRELIMAT'). 


\appendix

  \section{Time-dependent spin-wave theory }\label{app:spinwave}
  In this Section we provide further information on the time-dependent spin-wave expansion, referring the  reader to~\cite{jamir, jamir2} for  a comprehensive discussion on the method.

 We use the shorthand $O\equiv \langle \hat{O}\rangle$ for expectation values of  operators $\hat O$, and define the coordinates for the collective spin vector, $\vec{S}=\frac{N}{2}(\sin\theta\cos\phi,\sin\theta\sin\phi,\cos\theta)$, in terms of the polar, $\theta$, and azimuthal angle, $\phi$, on the Bloch sphere,  which allows writing compactly the  equations of motion:
\begin{subequations}\label{eq:a}
  \begin{align}
  \dot{\theta}&=-\frac{2\lambda(t)}{\sqrt{N}}(a+a^*)\sin\phi-J(1-\epsilon)\sin\theta\sin\phi\cos\phi\nonumber\\
  &+J\delta_{pp}\sin\theta\sin\phi\cos\phi-J\delta_{pq}\sin\theta\cos\theta\cos^2\phi,\\
  \dot{\phi}&=\omega_0-\frac{2\lambda(t)}{\sqrt{N}}(a+a^*)\cot\theta\cos\phi-J(1-\epsilon)\cos\theta\cos^2\!\phi \nonumber\\
  &+J\delta_{qq}\cos\theta\cos^2\phi-4J\delta_{pq}\sin\phi\cos\phi,\\
    \dot{a}&=-i\omega a-\frac{\kappa}{2}a-i\lambda(t)\sqrt{N}(1-\epsilon)\sin\theta\cos\phi.
\end{align}
\end{subequations}
The effect of $H_{int}$ enters the dynamics of the collective spin vector via the spin-wave correlations $\delta_{\alpha\beta} (\alpha, \beta = q,p)$ and the spin-wave density $\epsilon$, which are in turn dynamically coupled to the collective spin and photon field. Here,
\begin{eqnarray}
      \delta_{\alpha\beta}&=&\frac{2}{N}\sum_{k\neq 0}\cos k\Delta_k^{\alpha\beta},~~ \text{and}\\
      \epsilon&=&\frac{1}{N}\sum_{k\neq 0}(\Delta_k^{pp}+\Delta_k^{qq}-1),
\end{eqnarray}
with 
    \begin{eqnarray}
    \Delta_k^{\alpha\alpha}&=&\langle \alpha_k(t)\alpha_{-k}(t)\rangle, ~{\rm with}~\alpha=q,p,\\
    \Delta_k^{pq}&=&\frac{1}{2}(\langle p_k(t)q_{-k}(t)\rangle+\langle q_k(t)p_{-k}(t)\rangle),                    
    \end{eqnarray}
where $q_k$ and $p_k$  are the  canonically conjugated bosonic variables associated with spin-waves with wave-vector $k\neq0$.
     
Following the procedure in \cite{jamir2}, we derive the equations of motion for the spin-wave correlations
\begin{equation}\label{eq:sw}
\begin{split}
  \dot{\Delta}_k^{qq}&=-\frac{4\lambda(t)}{\sqrt{N}}(a+a^*)\frac{\cos\phi}{\sin\theta}\Delta_k^{pq}\\
  &-2J(\cos^2\phi-\cos k\sin^2\phi)\Delta_k^{pq}\\
  &-2J\cos k\cos\theta\sin\phi\cos\phi\Delta_k^{qq},\\
  \dot{\Delta}_k^{pp}&=\frac{4\lambda(t)}{\sqrt{N}}(a+a^*)\frac{\cos\phi}{\sin\theta}\Delta_k^{pq}\\
  & +2J(\cos^2\phi-\cos k\cos^2\theta\cos^2\phi)\Delta_k^{pq}\\&+2J\cos k\cos\theta\sin\phi\cos\phi\Delta_k^{pp},\\
  \dot{\Delta}_k^{pq}&=\frac{2\lambda(t)}{\sqrt{N}}(a+a^*)\frac{\cos\phi}{\sin\theta}(\Delta_k^{qq}-\Delta_k^{pp})\\
  &+J(\cos^2\phi-\cos k\cos^2\theta\cos^2\phi)\Delta_k^{qq}\\
  &-J(\cos^2\phi-\cos k\sin^2\phi)\Delta_k^{pp}.
\end{split}
\end{equation}
These quantities intertwine with the equations of motion  \eqref{eq:a}, and represent the feedback of the non-equilibrium Gaussian fluctuations of spin waves on the motion of the collective spin (the $k=0$ mode). The self-consistent solution of Eqs.~\eqref{eq:a} and~\eqref{eq:sw} yields the dynamics of the model. 
We notice that these equations of motion  are derived in the thermodynamic limit~\cite{jamir2}, therefore we expect our results to be  insensitive to the choice of $N$, as we checked for the main results of our work.

In the last equation of Eqs.~(\ref{eq:a}), the phase fluctuation accompanied with photon loss has been neglected. This is valid in the large $N$ limit, since the time crystals under scrutiny here exist in the superradiant regime, where the photon number is $\propto N$, and thus photon noises become subleading. For small $N$, the effect of phase fluctuation can be incorporated in the spin-wave approach by adding a Langevin noise term in Eqs.~\ref{eq:a}(c).

\begin{figure}[t!]
  \includegraphics[width=0.3\textwidth]{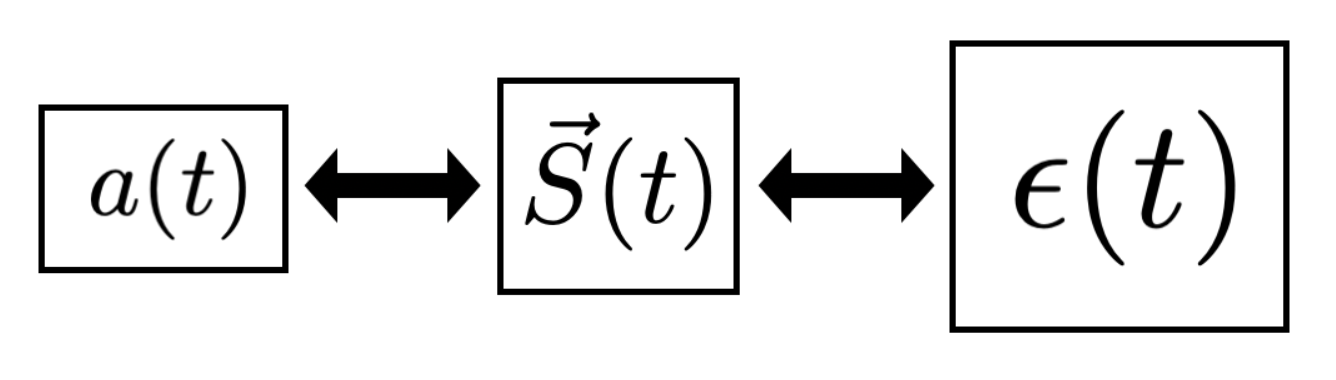}
  \caption{In the time-dependent spin wave theory, quantum fluctuations introduced by the short-range interaction term $H_{int}$ are included as a self-generated bath which  couples to the order parameter, $\vec{S}(t)$. The latter  couples also to the photonic mode which is cooled by a zero temperature bath. Therefore, the dynamics of the order parameter results from the competing interactions with an internal spin wave bath (represented by spin waves' density, $\epsilon(t)$,  in the sketch above), and with an external cold bath, mediated by the cavity photon, $a(t)$.}
\label{figmet}
\end{figure}

The spin wave density $\epsilon$ has to remain small during the course of the evolution, in order to render consistent the lowest order Holstein-Primakoff expansion, employed to derive Eqs.~\eqref{eq:a} and~\eqref{eq:sw}. This restriction of the spin-wave approach limits resolving possible  structures in the dynamical responses when the heating from $J$ is significant, which might be interesting to explore in future works. For $J=0$, one can readily see from the above equations that $\epsilon(t)=0$ at any time, while for $J\neq0$, the   length of the collective spin is shrinked via $2|\vec{S}(t)|/N=1-\epsilon(t)$.  
A pictorial description of the method is provided in Fig.~\ref{figmet}. 

\section{Persistence of Dicke-TC}\label{app:S1}
\begin{figure}[t!]
  \includegraphics[width=0.35\textwidth]{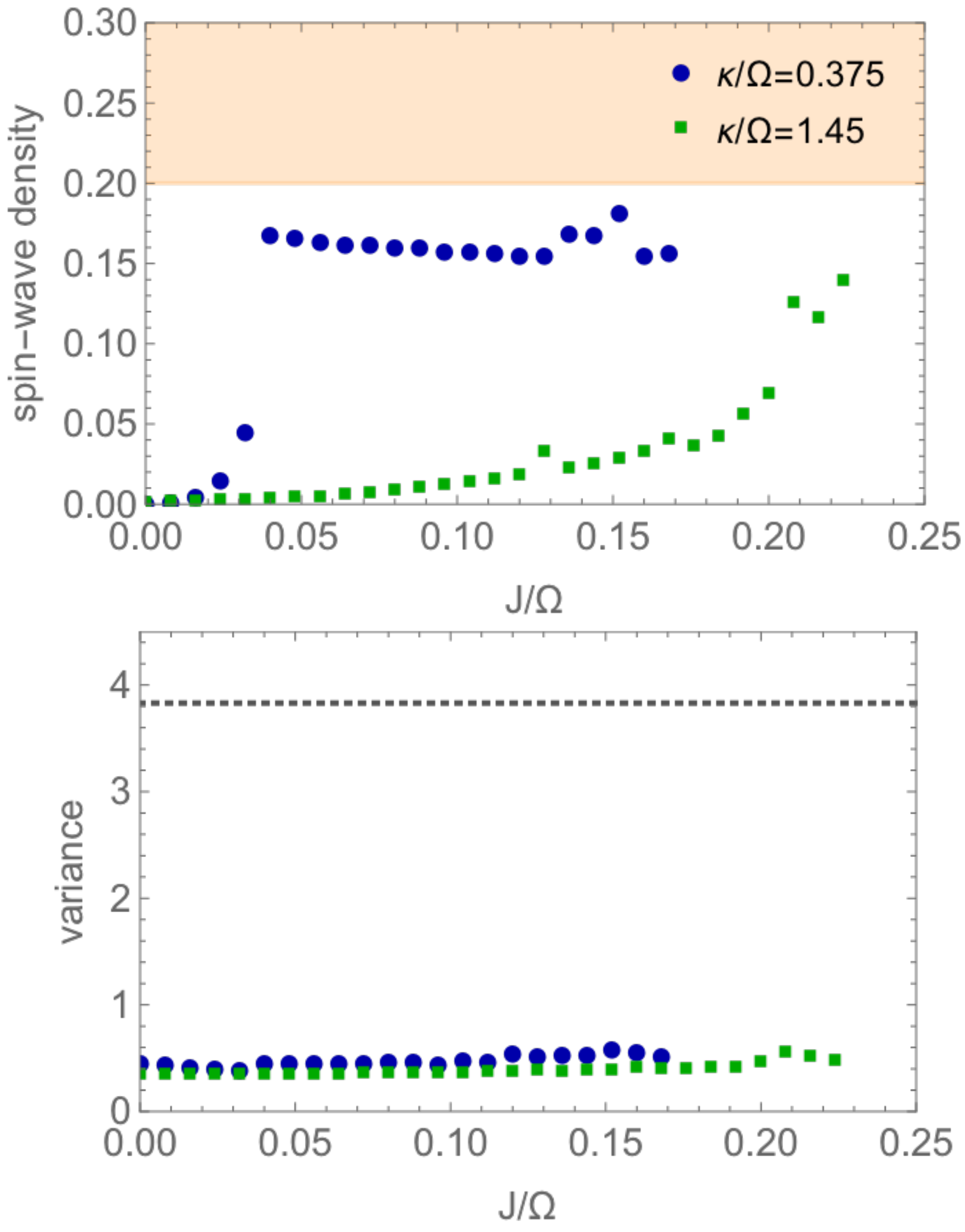}
  \caption{The maximum spin-wave density $\epsilon(t)$ within 5000 periods of dynamics (top) and the variance of Fourier spectrum (bottom) for various interaction strengths $J$ at fixed dissipation rate $\kappa$ corresponding to the Dicke-TC shown in Fig.~\ref{fig2}(a) (blue dotes), and (b) (green squares) and $\delta=-0.12$. The variance is not calculated for those values of $J$ corresponding to  spin-wave density  above the threshold (orange region). As a reference, the dotted line in the bottom panel shows the variance obtained for the IR dynamics in Fig.~\ref{fig2}(c). }
  \label{figS1}
  \end{figure}

In this section we provide more details of the Dicke-TC behaviour discussed in Sec.~III. In the top panel of Fig.~\ref{figS1}, we plot the maximum  spin-wave density over the 5000 periods of dynamics obtained from the above time-dependent spin-wave approach at stroboscopic times, when increasing the interaction strength $J$. As discussed in the main text, when the system is prone to many-body heating, sizable amount of spin-waves develops, and the Holstein-Primakoff expansion breaks down. Here, we choose as upper threshold for spin-wave density at the value, $\epsilon\sim 0.2$. From the spin dynamics, we can also calculate the variance $\chi$ of the Fourier spectrum of $S_x+iS_y$, defined as $\chi=\int_{\nu>0} d\nu p(\nu) (\nu-0.5\Omega)/\int_{\nu>0} d\nu p(\nu)+\int_{\nu<0} d\nu p(\nu) (\nu+0.5\Omega)/\int_{\nu<0} d\nu p(\nu)$, where $p(\nu)$ is the height of the Fourier spectrum at frequency $\nu$. TCs feature stable subharmonic responses, with the spectrum composed of two sharp peaks located at $\nu=\pm0.5\Omega$, and thus a small value of $\chi$.  The variance for increasing $J$ is plotted in the bottom panel of Fig.~\ref{figS1}. With a small $\kappa/\Omega<1$, the spin-wave density quickly grows with $J$, which is reduced  with a large $\kappa/\Omega>1$. The variance $\chi$ remains negligible for values of $J$ with low spin-wave density, indicating the existence of the Dicke-TC with finite $J$.  

\begin{figure}[t!]
  \includegraphics[width=0.45\textwidth]{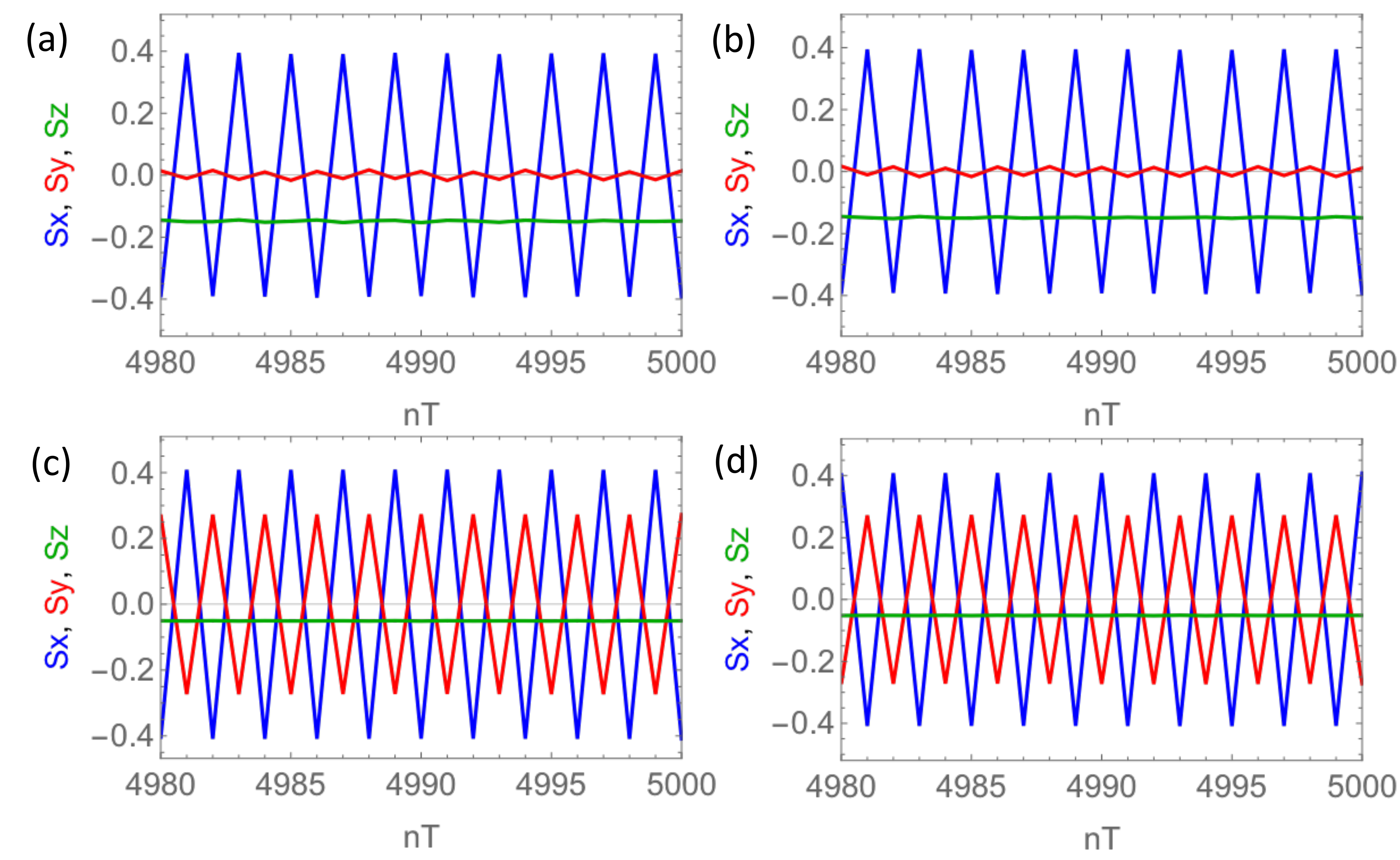}
  \caption{Dicke TCs for different initial conditions, with the same parameters as in Fig.~2a (upper panels), and Fig.~2b (lower panels). The initial state in panels (a) and (c) is a collective spin state with $\theta=0.5\pi$ and $\phi=0.3\pi$ (see notation in Appendix A), while the initial state in panels (b) and (d) is characterised by $\theta=0.3\pi$ and $\phi=0.3\pi$. The system exhibits similar Dicke-TC order for these different initial states. }
\label{init}
\end{figure}

In Fig.~\ref{fig2} an initial state with $\theta=0.5\pi$ and $\phi=0$ has been used. We have also checked in our numerical simulations that the Dicke-TC is insensitive to initial conditions, thanks to the dissipative nature of dynamics: we observe similar sub-harmonic dynamics over a wide range of different initial states, as demonstrated in Fig.~\ref{init}.

\section{Mean-field analysis of dynamical responses}
\begin{figure}[t]
\includegraphics[width=0.3\textwidth]{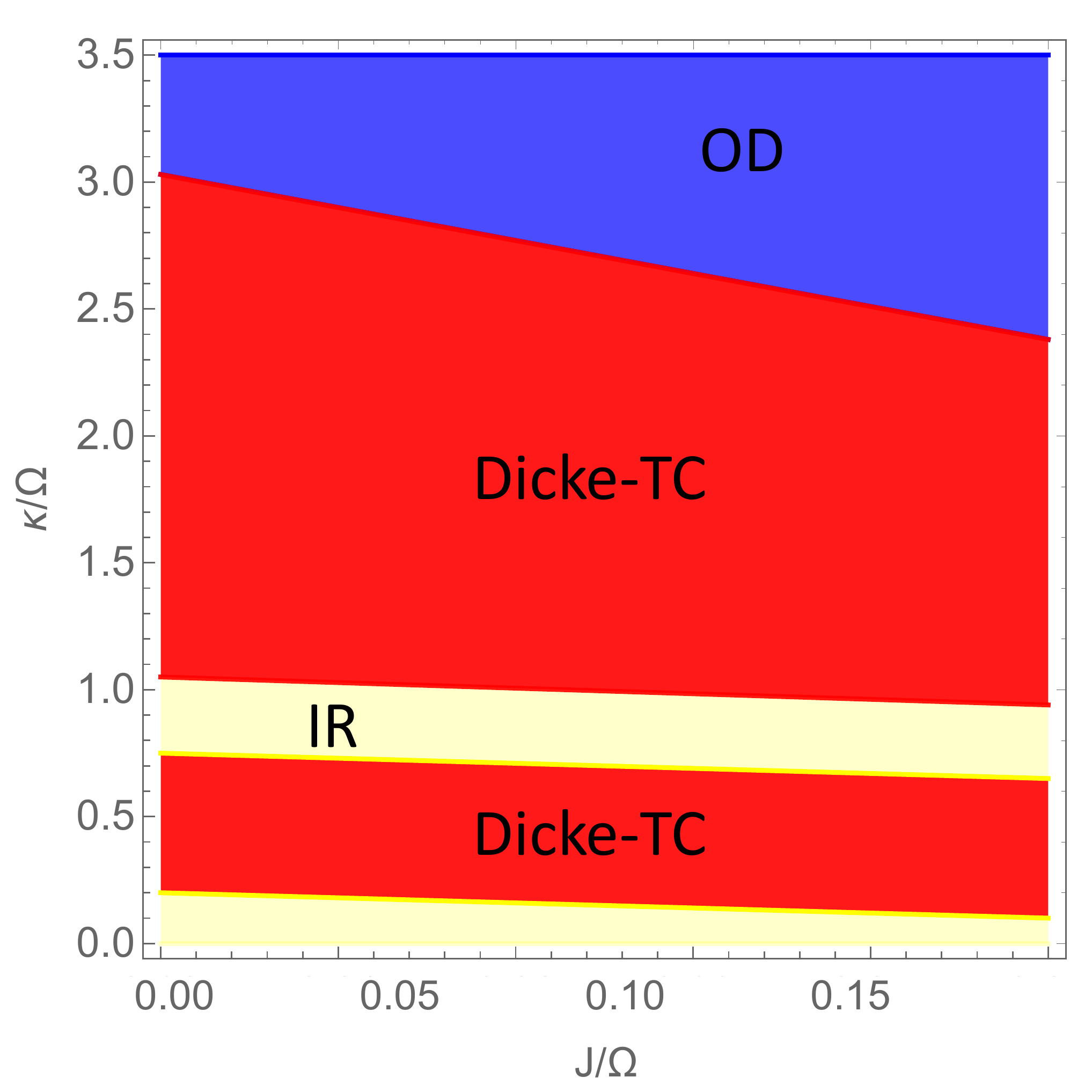}
\caption{Mean-field dynamical responses for $\delta=-0.12$, as a function of dissipation rate, $\kappa$, and many-body interactions strength, $J$.}
\label{figMF}
\end{figure}

To understand the dynamical responses in our driven-dissipative many-body system, here, we first apply a mean-field treatment to solve dynamics from the master equation Eq.~(3). In Fig.~\ref{figMF} we display the various dynamical responses of the system. The  boundaries between the IR  and the Dicke-TC regions are  dictated by a transition in  the variance of the Fourier spectrum of $ S_x(t)$, which is negligible  in the case of the Dicke-TC. The OD regime is identified with a vanishing  $|S_x+iS_y|$ at $5000$ periods, which exhibits a second-order phase transition when crossing into the Dicke-TC regime. In a mean-field analysis, quantum fluctuations are neglected and thus 
it  does not predict  the MTC,  the heating region, and the enhanced robustness for ferromagnetic Dicke-TC, which are the genuine many-body results of our study. Instead, it shows as an artefact the persistence of Dicke-TC despite strong interactions.

\section{Summary of dynamical responses from time-dependent spin-wave analysis}\label{app:phase}
To account for quantum many-body effects, we solve Eqs.~\eqref{eq:a} to~\eqref{eq:sw} in the time-dependent spin-wave approach to obtain the spin dynamics. We explore over a range of parameters $\kappa$ and $J$ and find rich dynamical responses. Fig.~\ref{figsketch} shows a qualitative sketch of regions where different dynamical behaviours are observed. We label the region with large spin-wave density $\epsilon(t)\geq 0.2$ as heating (H), where the spin-wave treatment breaks down.  IR regime is characterised with a large variance $\chi$ of the Fourier spectrum together with $\epsilon(t)<0.2$. As discussed in Ref.~\cite{Gong} for the case of $J=0$, chaotic dynamics may arise in the collective Dicke model, which can result in large numerical errors. Here, we also find that when $\kappa/\Omega\sim 1$ the numerical integration tends to be unstable, and we associate these instances with IR as well in Fig.~\ref{figsketch}. The MTC region is identified with a slow decay of $S_x$ to a nonzero value at 5000 periods, with a small $\chi$ and $\epsilon(t)$ remaining below $0.2$. The boundary between the OD and the Dicke-TC regime resembles the one in a mean-field analysis, except that at a small value of $J$ it is interrupted by the emergence of other dynamical behaviours and heating: for increasing $J$, the boundary is set by a smaller, rather than a larger $\kappa$, suggesting that faster dissipation does not always protects the system against the heating from many-body interactions. As noted in Appendix \ref{app:spinwave}, limited by the choice of truncation in $\epsilon(t)$, a quantitative identification of the parameter regime for all dynamical responses and the nature of their boundaries is beyond the scope of this work.

\section{Comparison with integrability breaking  of the periodically kicked 'LMG' model }\label{app:lmg}
 In order to exemplify the non-trivial interplay that dissipation can have with many-body interactions, we have considered a similar analysis in a case governed by purely unitary dynamics. We have studied the Lipkin-Meshkov-Glick (LMG) model perturbed by $H_J$, 
\begin{equation}\label{LMG}
\begin{split}
H'=H_{(LMG)}+\hat{H}_J,~\text{with}\quad
H_{(LMG)}=-\frac{\lambda}{N}\hat{S}^2_x-g\hat{S}_z.
\end{split}
\end{equation}  
The dynamics entailed by $H'$ is periodically perturbed  by a collective rotation along the $\hat{x}$-axis;
the evolution operator reads in a period
\begin{equation}\label{kick}
\hat{U}=\hat{U}_{kick}\exp\left[-i\hat{H}_{(LMG)}~T\right],
\end{equation}
with $\hat{U}_{kick}\equiv \exp\left[-i\phi S_x\right]$.
This protocol has been shown in Ref.~\cite{RussomannoF} to display  TC behaviour when $\phi=\pi$  (for instance, in the dynamics of the stroboscopic transverse magnetization). The TC in this case is robust to displacements around the $\phi=\pi$ point, i.e. for angles $\phi=\pi\pm \delta$ (with $\delta>0$). We have  chosen this system as a comparison for the Dicke dynamics studied in the main text, since the latter effectively reduces  to the LMG model via adiabatic elimination of the photon mode for large cavity detunings.

\begin{figure}[t!]
\includegraphics[width=0.45\textwidth]{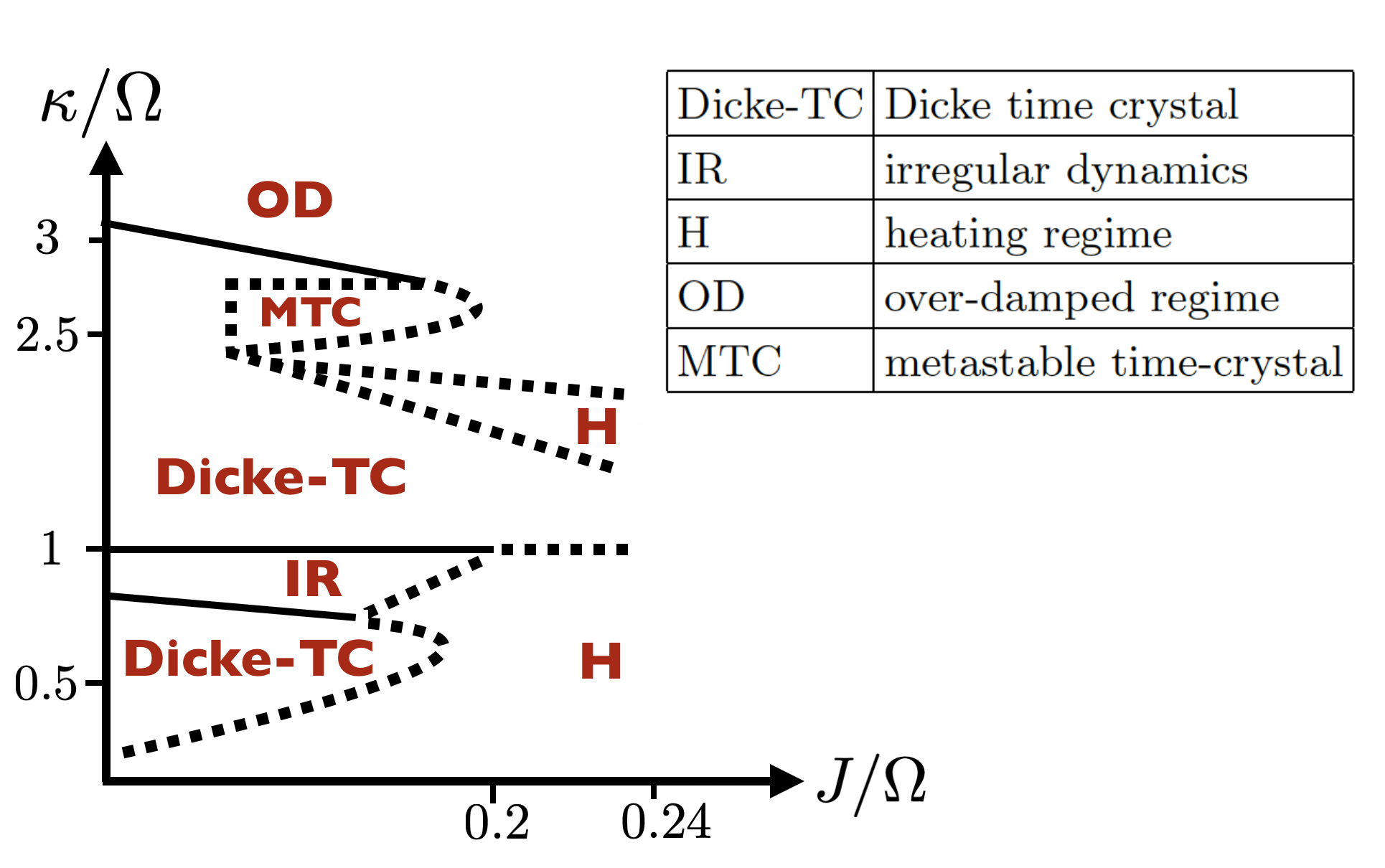}
\caption{A qualitative cartoon of the dynamical responses in the driven-dissipative many-body Dicke model, for varied dissipation rate, $\kappa/\Omega$, and many-body interaction strength, $J>0$, with $\delta = -ˆ'0.12$.  Lines separate regimes with qualitatively different dynamical responses; dotted lines at the boundary with the heating (H) regime indicate a crossover into regions where the collective spin order decays and the system becomes prone to many-body heating. The DTC responses are stable in the long-time limit, while the MTC represents a slowly decaying dynamical response, with vanishing dynamical order at long times. The OD region describes a regime dominated by dissipation where the order parameter quickly drops to zero.}
\label{figsketch}
\end{figure}

The periodically driven unitary dynamics in~\eqref{kick} does not entail a rich set of dynamical responses as  in Fig.~\ref{figsketch}: the TC persists (most likely in a pre-thermal fashion) for values of $J$ smaller than a certain critical threshold, $J_c$, above  which the system develops sizeable spin-wave density and therefore crosses over into a regime of 'heating'. This suggests that the presence of dissipation enriches the dynamical responses of a periodically driven interacting quantum many-body system.

  \section{ Quadruple-period dynamical response }\label{app:quad}

  We consider the Hamiltonian 
  \begin{equation}
\hat H=\omega \hat a^\dagger\hat a+\frac{\omega_0}{2}\sum_i\hat\sigma_i^z+\frac{\lambda}{N}(\hat a^\dagger+\hat a)\sum_i[e^{j{\bf k}\cdot {\bf x}_i}\hat\sigma_i^++h.c.],\label{eq:effH}
    \end{equation}
    with the  coupling strength between atoms and cavity   controlled by tuning the  phase factor $\propto{\bf k}\cdot {\bf x}_i$; such coupling can be realized by choosing proper laser dressing in a quantum optics setup. When  $e^{j{\bf k}\cdot{\bf x}_i}= 1$, the hamiltonian~\eqref{eq:effH} becomes  the conventional  Dicke model. In the case of $e^{j{\bf k}\cdot{\bf x}_i}\neq  1$, the coupling strength varies from site to site. We choose  $e^{j{\bf k}\cdot{\bf x}_i}=(-1)^{i/2}$, therefore, the hamiltonian~(\ref{eq:effH}) becomes
\begin{eqnarray}
  \hat H&=&\omega \hat a^\dagger\hat a+\frac{\omega_0}{2}\sum_i\hat\sigma_i^z+\frac{\lambda}{\sqrt{N}}(\hat a^\dagger+\hat a)\nonumber\\
            &&\times\sum_i[\hat\sigma_i^x\cos(i\pi/2) -\hat\sigma_i^y\sin(i\pi/2)].\label{eq:Hac}
\end{eqnarray}

\begin{figure}[t!]
  \centering
 \includegraphics[width=0.3\textwidth]{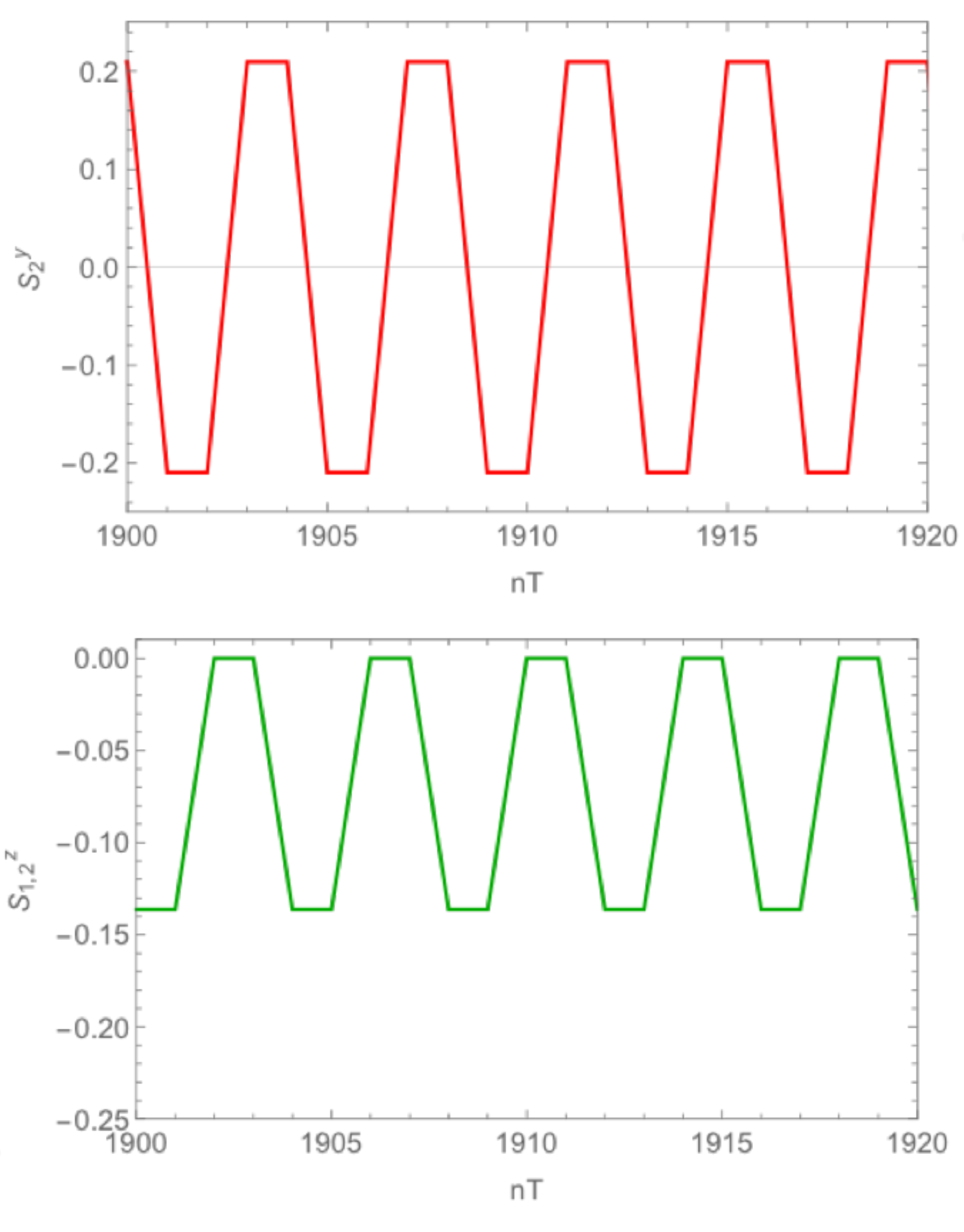}
 \caption{Stroboscopic  dynamics (top panel)  of $\sum_iS_i^y$ (for $i$ in every second site), and (bottom pannel) of  $\sum_iS_i^z$ (for $i$ in sublattices A and B), displaying quadruple response. }\label{fig:quadperf}
\end{figure}

We divide the system into four sublattices, A, B, C and D, with ${\rm mod}[i,4]=\{0,1,2,3\}$, respectively. Within each sublattice, atoms are collectively coupled to cavity photons. In the limit of $N\rightarrow\infty$, the stationary solution of the corresponding master equation hosts multiple possible  steady states, depending on the value $Q=\frac{\lambda^2\omega/\omega_0}{\kappa^2/4+\omega^2}$. The case $Q<1/4$ corresponds to the normal state without superradiance.  When $1/4\leq Q\leq 1/2$, the system is superradiant, with photon occupation
\begin{eqnarray}
  n_{cav}&=&\frac{N\omega_0(Q^2-1/16)}{\omega Q} ~~~\text{(group I)},
\end{eqnarray}
and similar to conventional Dicke superradiance, there are two possible spin states:
\begin{subequations}\label{eq:ss1}
\begin{eqnarray}
  \sigma_A^x&=&X,~\sigma_B^y=-X,~\sigma_C^x=-X,~\sigma_D^y=X,\\
  \sigma_A^x&=&-X,~\sigma_B^y=X,~\sigma_C^x=X,~\sigma_D^y=-X,
\end{eqnarray}
\end{subequations}
with $X=\sqrt{Q^2-1/16}/Q$, and both with $\sigma_A^z=\sigma_B^z=\sigma_C^z=\sigma_D^z=-\sqrt{1-X^2}$.
In this case the spin state at each site is in phase with the corresponding photon state. When $Q>1/2$, the steady state consists of two groups. The first group is the same as in~\eqref{eq:ss1}, while the  photon occupation in the second group is given by
  \begin{eqnarray}
   n_{cav}&=&\frac{N\omega_0(Q^2-1/4)}{4\omega Q} ~~~\text{(group II)}.
  \end{eqnarray}
  This group includes 8 different spin configurations
  \begin{subequations}\label{eq:ss2}
  \begin{eqnarray}
\sigma_A^x&=&X',\sigma_A^z=\!\!-Z',\sigma_B^y=X',\sigma_B^z=Z',\sigma_C^x=-X',\nonumber\\&&\sigma_C^z=-Z',\sigma_D^y=X',\sigma_D^z=-Z',\\
  \sigma_A^x&=&X',\sigma_A^z=\!\!-Z',\sigma_B^y=\!\!-X',\sigma_B^z=\!\!-Z',\sigma_C^x=-X',\nonumber\\&&\sigma_C^z=\!\!-Z',\sigma_D^y=-X',\sigma_D^z=Z',\\
    \sigma_A^x&=&-X',\sigma_A^z=\!\!-Z',\sigma_B^y=X',\sigma_B^z=\!\!-Z',\sigma_C^x=X',\nonumber\\&&\sigma_C^z=\!\!-Z',\sigma_D^y=X',\sigma_D^z=Z',\\
    \sigma_A^x&=&\!\!-X',\sigma_A^z=\!\!-Z',\sigma_B^y=-X',\sigma_B^z=Z',\sigma_C^x=X',\nonumber\\&&\sigma_C^z=\!\!-Z',\sigma_D^y=-X',\sigma_D^z=\!\!-Z',\\
    \sigma_A^x&=&X',\sigma_A^z=Z',\sigma_B^y=X',\sigma_B^z=\!\!-Z',\sigma_C^x=X',\nonumber\\&&\sigma_C^z=\!\!-Z',\sigma_D^y=-X',\sigma_D^z=\!\!-Z',\\
  \sigma_A^x&=&X',\sigma_A^z=\!\!-Z',\sigma_B^y=-X',\sigma_B^z=\!\!-Z',\sigma_C^x=X',\nonumber\\&&\sigma_C^z=Z',\sigma_D^y=X',\sigma_D^z=\!\!-Z',\\
    \sigma_A^x&=&\!\!-X',\sigma_A^z=\!\!-Z',\sigma_B^y=X',\sigma_B^z=\!\!-Z',\sigma_C^x=-X',\nonumber\\&&\sigma_C^z=Z',\sigma_D^y=-X',\sigma_D^z=\!\!-Z',\\
    \sigma_A^x&=&\!\!-X',\sigma_A^z=Z',\sigma_B^y=-X',\sigma_B^z=\!\!-Z',\sigma_C^x=-X',\nonumber\\&&\sigma_C^z=\!\!-Z',\sigma_D^y=X',\sigma_D^z=\!\!-Z',
  \end{eqnarray}
  \end{subequations}
  with $X'=\sqrt{Q^2-1/4}/Q$, and $Z'=\sqrt{1-X'^2}$. These states correspond to having a ``defect'' in the spin configurations, and thus results in a lower photon number. A linear stability analysis suggests that the above  states can all be stable. The existence of multiple  steady states provides the possibility of producing subharmonic responses to external driving. 

When $1/4\leq Q\leq 1/2$, we can have a period-doubled  dynamical response if we apply a Floquet driving scheme similar to the one discussed in the main text.

When $Q>1/2$, in addition to period doubling, we can have a dynamics with period $4T$. However, this would require driving that can convert one steady state  to another in Eq.~(\ref{eq:ss2}). A possible procedure is as follows: We initialise  spins close to $\sigma_A^x=\sigma_B^y=-\sigma_C^x=\sigma_D^y=1$, and let the atom-cavity system interact for some time to reach steady state, which can be diagnosed via monitoring the photon emission from cavity. Then, we start to apply a driving pulse $\tilde\theta$   at the end of each period $T$. $\tilde\theta$ consists of single-site rotations. Specifically, we apply $\hat R_{z}(\pi)$ rotation to all odd sites (A and C); even sites (B and D) are rotated depending on the measurement outcome of $\sum_{i\in A}S_i^x$ prior to the pulse: $\hat R_{z}(\pi)$ for negative outcome, and $\hat R_{y}(\pi)$ for positive outcome. The pulse strength is kept equal to cavity detuning. 

If we  measure $\sum_iS_i^y$ for $i$ in all even sites, we will observe a quadruple period in dynamics, as plotted in Fig.~\ref{fig:quadperf}, demonstrating that the system indeed undergoes several steady states during the Floquet dynamics.  Including small photon loss during $\tilde\theta$, we still see oscillations at a stable $4T$ period. 
An alternative order parameter  is the inversion $\sum_iS_i^z$  in sublattices A and B, the dynamics of which also exhibits a period $4T$ (see again Fig.~\ref{fig:quadperf}). 

The above analysis assumed that atoms within each sub-lattice remain in the collective manifold and that therefore there are no quantum correlations. When atom number is sufficiently large, this represents a reliable approximation; notice, however, that even for a small  number of atoms, period doubling has been observed in Ref.~\cite{Gong} to persist for times longer than the decay time, therefore we can expect similar conclusions to hold for the quadruple  period dynamics discussed here.

\bibliography{bibliotimecrystal}

\end{document}